\documentclass[prc,twocolumn,floatfix,nofootinbib,superscriptaddress,
longbibliography]{revtex4-1}

\usepackage{amsmath}
\usepackage{amssymb}
\usepackage[utf8]{inputenc}
\usepackage{color}
\definecolor{goodgreen}{rgb}{0.1,0.5,0}
\definecolor{goodred}{rgb}{0.7,0,0}
\definecolor{goodblue}{rgb}{0,0,0.8}
\usepackage[colorlinks,urlcolor=goodgreen,citecolor=blue,linkcolor=goodred]%
{hyperref}
\usepackage{graphicx}
\usepackage[version=4]{mhchem}
\usepackage{siunitx}

\newcommand{\Vg}{\ensuremath{V_\text{g}}}
\newcommand{\Vsd}{\ensuremath{V_\text{sd}}}

\begin{document}

\title{Quartz tuning-fork based carbon nanotube transfer into quantum device
geometries}

\author{S. Blien}
\author{P. Steger}
\author{A. Albang}
\author{N. Paradiso}
\author{A. K. Hüttel}
\email{andreas.huettel@ur.de}
\affiliation{Institute for Experimental and Applied Physics,
University of Regensburg, Universit\"{a}tsstr. 31, 93053 Regensburg, Germany}

\date{16 May 2018}

\begin{abstract}
With the objective of integrating single clean, as-grown carbon nanotubes
into complex circuits, we have developed a technique to grow nanotubes directly
on commercially available quartz tuning forks using a high temperature CVD
process. Multiple straight and aligned nanotubes bridge the
$>\SI{100}{\micro\metre}$ gap between the two tips.
The nanotubes are then lowered onto contact electrodes, electronically
characterized in situ, and subsequently cut loose from the tuning fork using a
high current. First quantum transport measurements of the resulting devices at
cryogenic temperatures display Coulomb blockade characteristics.
\end{abstract}

\maketitle

\section{Introduction}

A fabrication technique that has led to many remarkable observations in quantum 
transport is the in-situ growth of carbon nanotubes onto pre-existing 
electrodes and trenches in between them \cite{nmat-cao:745}. Published results 
range from Coulomb blockade transport spectroscopy of unperturbed electronic 
systems \cite{nphys-deshpande:314,nature-kuemmeth:448,pecker:natphys2013,magda} 
all the way to high quality factor mechanical resonators and strong interaction 
between single electron tunneling and vibrational motion
\cite{highq,strongcoupling,science-lassagne:1107,pssb-huettel,kondocharge}. A 
natural limitation of this technique is that the electrode chip is exposed to 
the conditions of chemical vapour deposition (CVD) nanotube growth, typically 
$10-\SI{30}{\minute}$ in a gas mixture of hydrogen and methane at 
$800-\SI{1000}{\celsius}$ \cite{nature-kong:878}. Only few thin film materials 
survive this process, notably platinum-tungsten combinations 
\cite{nmat-cao:745,highq} and rhenium or rhenium-molybdenum alloys 
\cite{apl-singh:222601,brokensu4,remo,toscres}. Still, fabrication remains 
challenging and the integration of more sensitive circuit elements such as, 
e.g., Josephson junctions, quasi impossible.

The separation of growth and measurement chip provides a compelling alternative 
to in-situ growth of CNTs 
\cite{nl-wu:1032,nnano-pei:630,ncomm-ranjan:7165,waissman,pssb-gramich:2496}.
For the subsequent transfer of the nanotubes from one to the other, several 
approaches exist. While pressing growth surfaces directly onto the measurement 
chip to transfer CNTs potentially provides many viable devices per fabrication 
step and allows the lithographic selection of suitable CNTs on the target 
surface for contacting \cite{Desjardins2017,nicola}, the integration of clean,
suspended CNTs into complex, large-scale circuits requires a controlled 
deposition of single macromolecules 
\cite{ncomm-ranjan:7165,waissman,pssb-gramich:2496}.

Here, we present a technique to grow clean CNTs between the two prongs of 
commercially available quartz tuning forks and subsequently deposit them onto 
contact electrodes of arbitrary material. We demonstrate the details of the 
substrates, the transfer, and the cutting process and show first low 
temperature transport data.

\section{CNT growth on quartz tuning forks}\label{fork}

\begin{figure}[tp]
\includegraphics[width=\columnwidth]{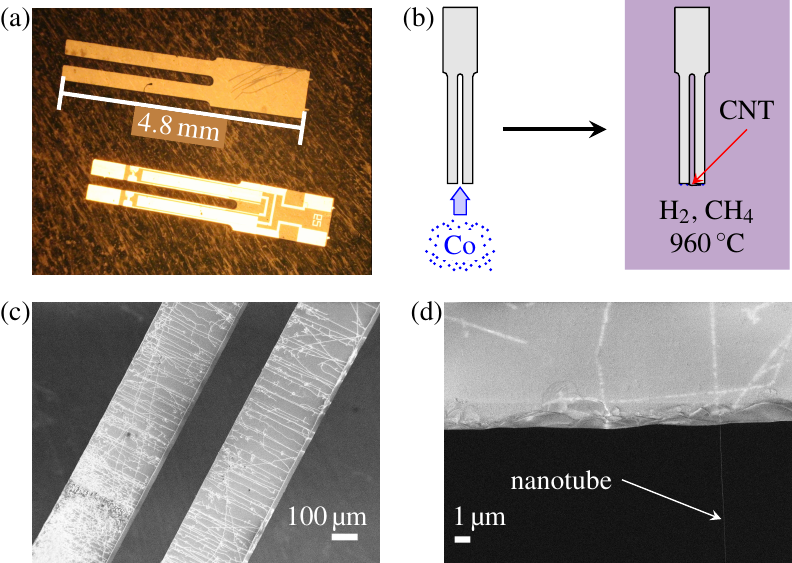} 
\caption{
(a) Commercial quartz tuning forks before and after removal of the 
metallization. 
(b) A thin Co layer is sputtered onto the tips of the fork as catalyst for the 
carbon nanotube growth by chemical vapour deposition. 
(c) Scanning electron micrograph of a fork after carbon nanotube growth: the 
nanotubes clearly display a preferred growth direction. For better visibility, 
here the entire fork surface has been covered with Co growth catalyst.
(d) Scanning electron micrograph of a carbon nanotube crossing the gap between 
the two fork tips.
}
\label{fig:fork}
\end{figure}

We start with a wafer piece containing several commercial-grade quartz tuning 
forks, see Fig.~\ref{fig:fork}(a). After breaking out one or more forks, the 
metallic contacts are removed using aqua regia, hot hydrochloric acid and hot 
NaOH baths and successive cleaning steps of sonication and plasma ashing. Then, 
a nominally \SI{1}{\nano\metre} thick layer of cobalt is sputter-deposited onto 
the tips of a fork, see Fig.~\ref{fig:fork}(b). For such a nominal thickness Co 
does not form a homogeneous film, but a randomly distributed ensemble of Co 
clusters which serve as catalyst centers for the carbon nanotube growth 
\cite{Kumar.2010,Yuan.2008}. 

As next step, the forks are placed on a glass plate and inserted into the 
quartz tube of a CVD furnace. The furnace is heated up under a steady flow of 
an argon / hydrogen mixture and then kept at \SI{960}{\celsius} for 30 minutes 
under a constant gas flow of methane and hydrogen. The flow rates, 
\SI{10}{sccm} \ce{CH4} and \SI{20}{sccm} \ce{H2}, are typical for clean CNT 
growth \cite{nature-kong:878}. The fork is placed perpendicular to the gas 
stream. As a result, the growth is directional in the sense that CNTs grow 
mainly in the prong-to-prong direction, see Fig.~\ref{fig:fork}(b) and also 
Fig.~\ref{fig:fork}(c,d), where the entire fork surface has been covered with 
catalyst for better visibility of the resulting nanotube growth. 

Imaging the forks in a scanning electron microscope after growth, we find that 
even with catalyst coating only the fork tips typically up to five nanotubes or 
nanotube bundles per fork are suspended over the gap between the tips 
\cite{Kasumov2007,magda}. To avoid damage and carbon contamination, we do not
image forks that are actually used for transfer. In a future setup one could
imagine using optical means, as, e.g., Raman or photoluminescence imaging
\cite{nl-lefebvre-2006} to count the suspended nanotubes between the fork
prongs.

\section{Target chip}

\begin{figure}[tp]
\includegraphics[width=\columnwidth]{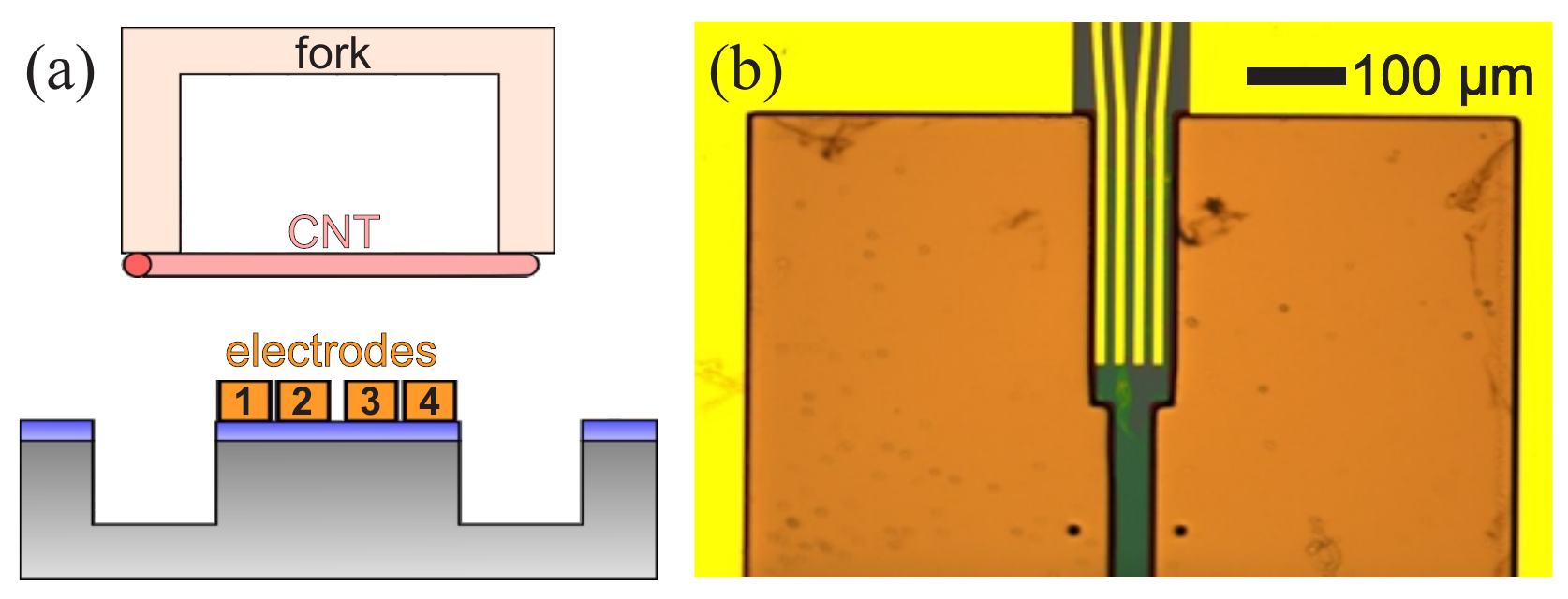} 
\caption{(a) Schematic of the carbon nanotube transfer: the fork carrying a 
nanotube is sunk into two trenches that are locally etched into a target chip 
on both sides of four gold electrodes. (b) Optical micrograph of the target
chip: four contact electrodes and a ground plane (yellow), the elevated center 
ridge carrying the electrodes (dark green), and surrounding deep-etched areas 
(orange) are visible.}
\label{fig:chip}
\end{figure}
For first tests of the transfer process, devices with four long electrodes were
prepared via optical lithography, see Fig.~\ref{fig:chip}(a) for a schematic 
side view and Fig.~\ref{fig:chip}(b) for a microscope top view. The substrate 
is highly p-doped silicon, with a \SI{500}{\nano\metre} thermally grown surface 
oxide. On its surface, four finger-like gold electrodes are deposited using 
thermal evaporation, and lift-off. The typical width of the electrodes and the 
distance between them are both \SI{10}{\micro\metre} for this simplified test
device. Next to the electrodes, two rectangular areas are locally etched to a 
depth of \SI{12}{\micro\metre} by an anisotropic reactive ion etching process 
using \ce{SF6} and Ar. The etch depth should be as large as possible and is 
mainly limited by the lithographic resist protecting the remaining structure.

\section{Transfer and cutting process}\label{transfer}

\begin{figure}[tp]
\includegraphics[width=\columnwidth]{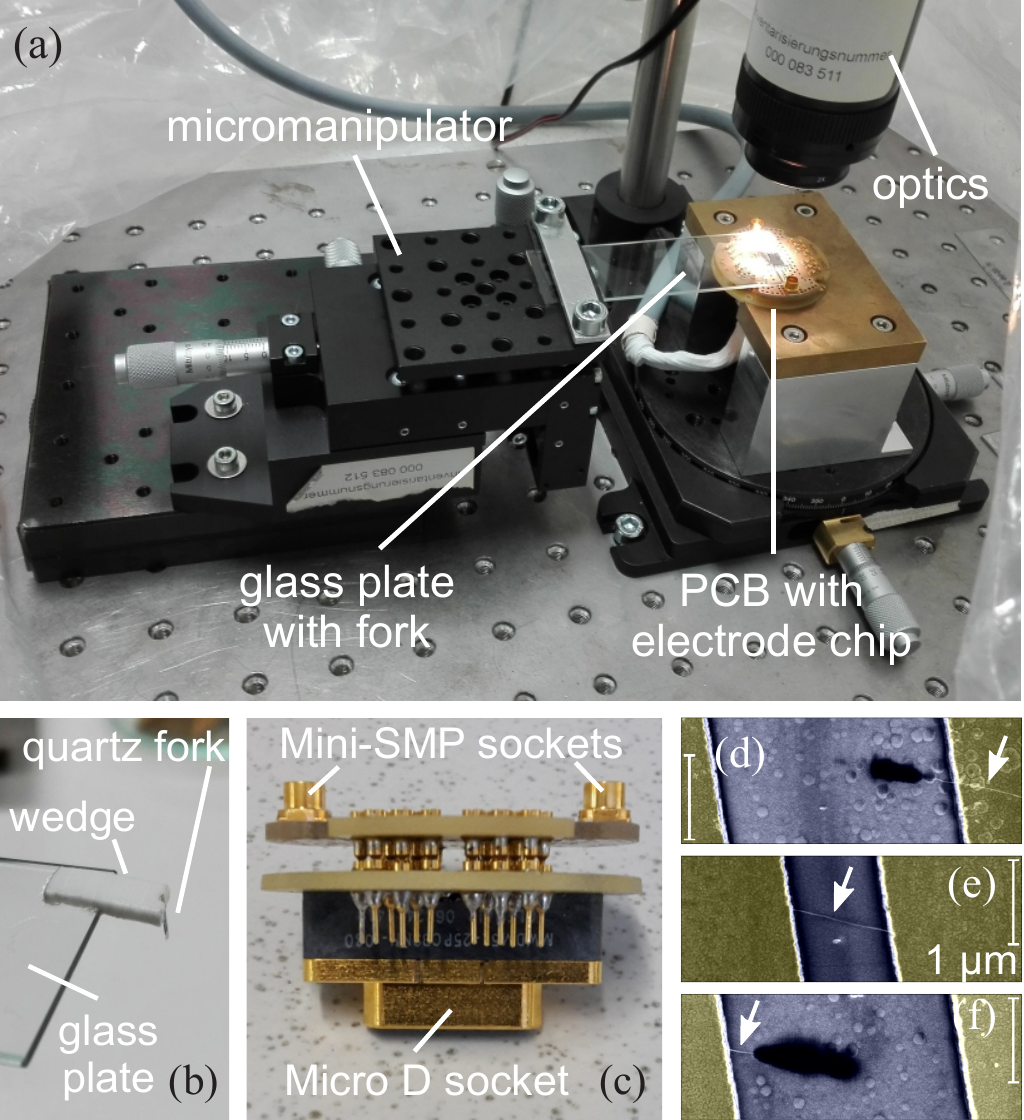}
\caption{
(a) Transfer setup: the quartz fork is mounted on a micromanipulator stage. It 
can be lowered to the target chip, which is glued onto a printed circuit board 
(PCB) and is electrically connected. The process is monitored via an optical 
microscope with a zoom lens and a camera. 
(b) Detail picture of how the quartz fork is mounted on the glass plate.
(c) Side view of the sample holder: to establish connection to electronic 
devices, a second PCB with a Micro D socket is attached. For further 
experiments, two high frequency ports with Mini-SMP connectors are additionally 
soldered on top of the board.
(d-f) Scanning electron micrographs of a successfully transferred CNT: the
nanotube has been cut between each pair of outer electrodes (d, f) and now only
connects the two inner electrodes (e).
}
\label{fig:transfer}
\end{figure}
For the transfer, the quartz fork carrying as-grown CNTs is attached to a 
glass object plate and mounted on a micromanipulator stage, see 
Figs.~\ref{fig:transfer}(a) and (b). The setup is adapted from the equipment 
combination used in \cite{castellanos-gomez} to dry-stamp 2D materials. As 
there, a camera combined with a zoom lens allows us to observe the target chip 
from the top. The base plate is modified insofar as it clamps a printed circuit 
board sample holder with a 25-pin MDM socket at the bottom, see 
Fig.~\ref{fig:transfer}(c). The target chip is glued onto the circuit board and 
bonded; the electrodes are electrically contacted during the transfer process.

Using the micromanipulator stage, the quartz fork is lowered onto the chip such 
that its tips sink into the deep-etched areas on both sides of the dc contacts, 
cf. Fig.~\ref{fig:chip}(a). The process is monitored both optically and 
electrically. On the one hand, we use the microscope camera to monitor the fork 
position during the alignment. On the other hand, by applying 100\,mV between 
contacts 1 and 4, see Fig.~\ref{fig:chip}(a), we can detect a CNT bridging the 
metal electrodes by simply measuring a finite current.
\begin{figure}[tp]
\includegraphics[width=\columnwidth]{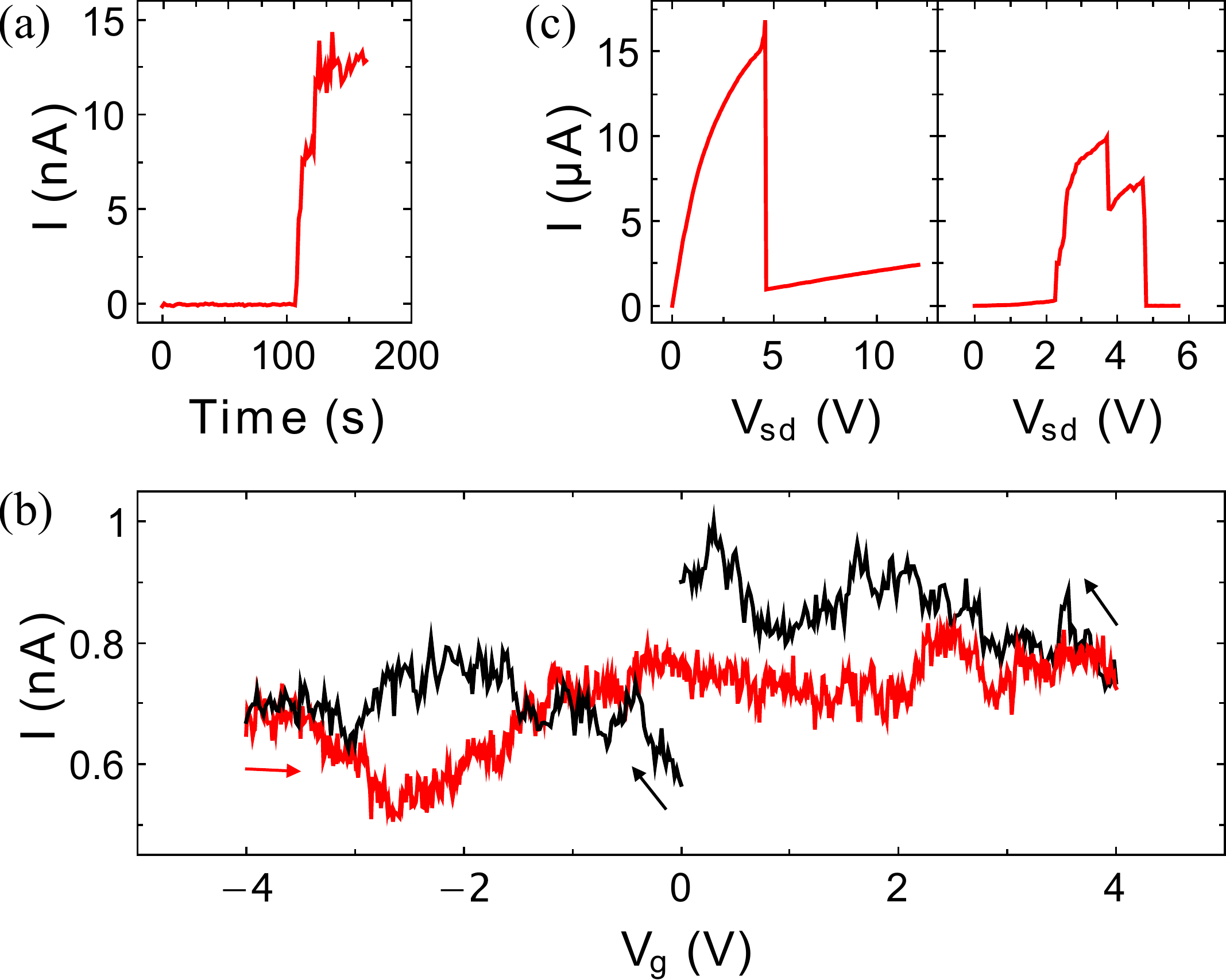}
\caption{
(a) The current between the voltage biased contacts 1 and 4, see
Fig.~\ref{fig:chip}(a), is measured continuously while a quartz fork is lowered 
onto the target chip. As soon as a CNT touches the electrodes a finite current 
can flow.
(b) Example back gate voltage sweep at a bias voltage of \SI{3}{\milli\volt}, 
recorded during a transfer process before cutting the nanotube. This allows 
estimating the type (metallic, semiconducting or bundle) of nanotube before 
finally leaving it on the device.
(c) Current measured during two different voltage ramps for ``cutting'' a CNT. 
From the shape of the resulting curves one can draw conclusions on the transfer 
result, see the text.
}
\label{fig:cutting}
\end{figure}
This is illustrated in Fig.~\ref{fig:cutting}(a), where at a time index of 
$t\approx 110\,\text{s}$ contact is made. Back gate voltage sweeps, see 
Fig.~\ref{fig:cutting}(b), then allow us to estimate whether a semiconducting 
or metallic nanotube or a nanotube bundle is contacted.

By ramping up a voltage bias and thereby the current between contacts 1 and 2, 
as well as subsequently between contacts 3 and 4, while the device is in air, 
the segments of the tube between these contacts can be electrically cut. 
Example current-voltage characteristics during this process are plotted in
Fig.~\ref{fig:cutting}(c). The critical current for cutting a nanotube 
typically lies in the range of $10-\SI{30}{\micro\ampere}$, consistent with the 
findings of Refs.~\cite{waissman,pssb-gramich:2496}. If at a certain point the 
current drops to zero in one single step as, e.g., in the left part of 
Fig.~\ref{fig:cutting}(c), this indicates that one single-wall carbon nanotube 
has been cut. If the current decreases to zero in several steps as in the right 
part of Fig.~\ref{fig:cutting}(c) the segment was a multi-wall nanotube or 
bundle and the steps correspond to breaking the shells or nanotubes one at a 
time. We were able to verify this interpretation of the number of steps in the 
I-V-curves by extracting the diameter of successfully transferred nanotubes 
from atomic force microscopy images at large contact distances, where the 
nanotubes can touch the substrate.

If the approach of fork and target chip is not done carefully enough, a nanotube
can be ripped off the fork tips and then fall down to the substrate in the 
deep-etched areas. Then, electrodes 1 and 2 are still electrically connected 
via the substrate even after the nanotube segment between them has been cut,  
resulting in a tail of finite current in the I-V-curve, cf. 
Fig.\ref{fig:cutting}(c), left panel. 

\section{Cleaning of the quartz forks for re-use}

After successful completion of the cutting process the detached nanotube lies 
only over the inner contact pair (2 and 3), as can be seen in the SEM image of 
Fig.~\ref{fig:transfer}(e). The quartz fork can then be safely lifted and
removed. 

Given the chemical and mechanical stability of the tuning forks, a rigorous 
cleaning procedure can subsequently be applied to remove both carbon residues 
and cobalt catalyst. We use plasma ashing to remove organic compounds grown in 
the preceding CVD process, and a bath of hot nitric acid to dissolve residues 
of old catalyst. After sonication and another short plasma ashing step the 
forks can be reintroduced into the fabrication cycle by sputtering a new layer
of Co catalyst.

\section{Low temperature characterization}

\begin{figure}[tp]
\includegraphics[width=\columnwidth]{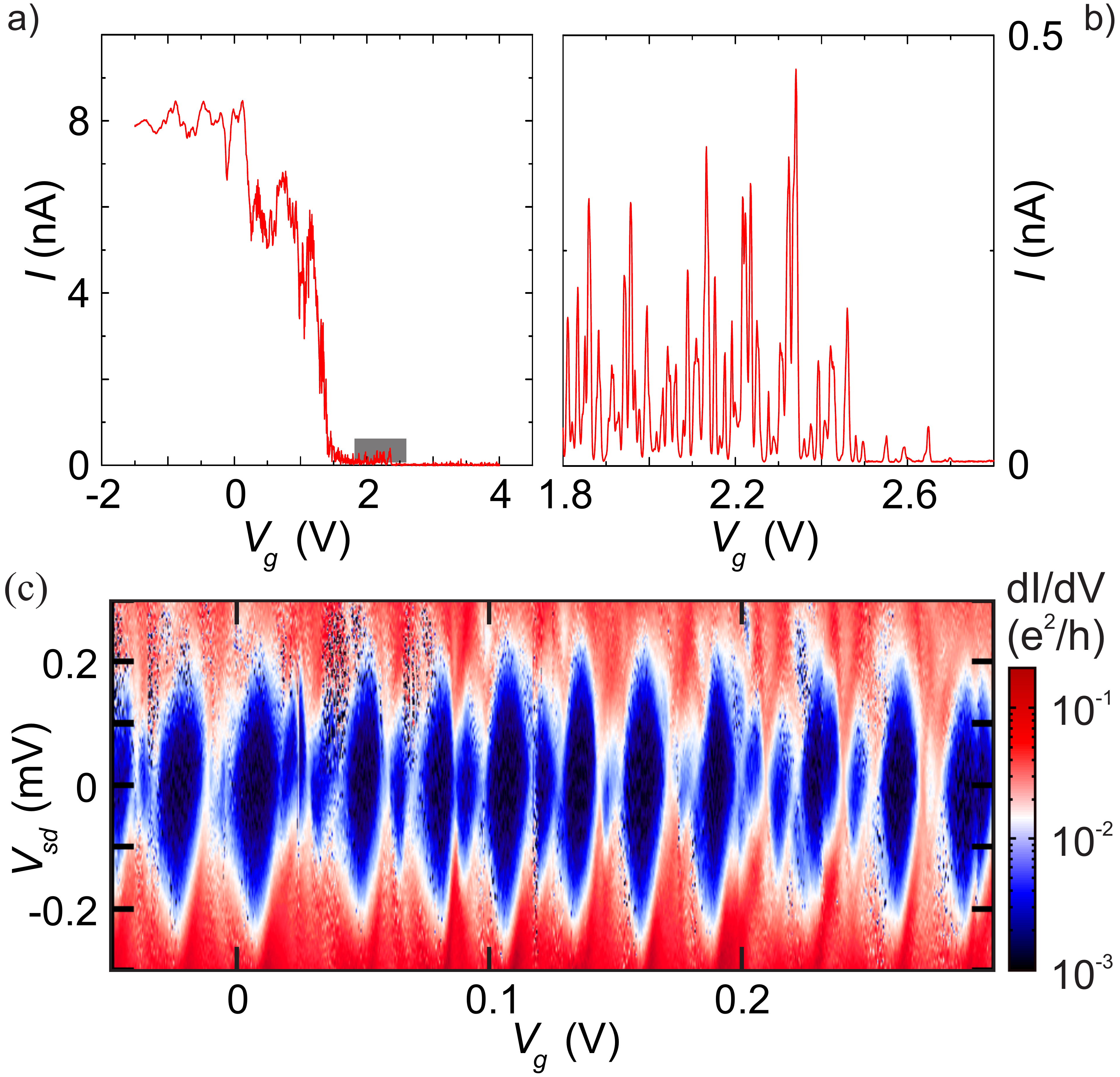} 
\caption{
(a) Characterization of a transferred CNT at $T=4.2 \,
\text{K}$. Plotted is the current through the nanotube as a function of the 
gate voltage \Vg, at an applied source-drain voltage of $\SI{2}{\milli\volt}$. 
Different parameter regions can be distinguished, see the text. 
(b) Zoom into the shaded area of (a), displaying Coulomb oscillations of the 
current. 
(c) Stability diagram of a transferred CNT at $T=\SI{15}{\milli\kelvin}$; 
differential conductance as function of gate voltage and source-drain voltage. 
A pattern of Coulomb blockade areas with two distinct sizes is visible. 
}
\label{fig:results}
\end{figure}

After successfully transferring a carbon nanotube to a substrate similar to the 
one shown in Fig.~\ref{fig:chip}, we have cooled down the device to
liquid helium temperature. The device was fabricated on a highly doped Si 
wafer, such that the substrate can be electrically connected and used as a 
global backgate. Fig.~\ref{fig:results}(a) shows the current through the CNT in 
dependence on the gate voltage \Vg, when 2\,mV bias is applied. Several 
distinct gate voltage regions can be distinguished in the figure. For 
$\Vg<1.8\,\text{V}$ the nanotube is strongly coupled to the electrodes, 
resulting in an open system. In the region $1.8\,\text{V}<\Vg<2.8\,\text{V}$ 
Coulomb blockade and single electron tunneling peaks are visible; see 
Fig.~\ref{fig:results}(b) for a detail zoom. For $2.8\,\text{V}<\Vg$ no current 
is flowing, indicating an electronic band gap.

A stability diagram at millikelvin temperatures of a similar device, where a 
carbon nanotube was deposited as described here, is shown in 
Fig.~\ref{fig:results}(c). The figure displays the differential conductance as 
function of the source-drain voltage \Vsd\ and a gate voltage \Vg. One can 
clearly identify the characteristic diamond pattern of Coulomb blockade regions 
as typically shown by quantum dots.

The stability diagram of Fig.~\ref{fig:results}(c) indicates a predominant 
electrostatic charging energy of approximately $E_c = 0.3\,\text{meV}$, 
corresponding to a total quantum dot capacitance of $C_\Sigma = {e^2}/{E_c} = 
530\,\text{aF}$. This is significantly larger than typical values for a device 
with single-wall nanotube length $l=1.4\,\mu\text{m}$ and a distance to the 
gate of $d=500\,\text{nm}$, the values expected from the contact geometry here. 
The small charging energy may indicate that multiwall nanotubes, bundles or 
nanotube networks have been transferred and measured. The appearance of an 
additional set of smaller Coulomb blockade areas in Fig.~\ref{fig:results}(c) 
supports this, indicating a second confined electronic system. No transversal 
mechanical resonance was found in transport measurements in a frequency range 
of $100\, \text{kHz} \leq f_\text{drive} \leq 500\,\text{MHz}$ \cite{highq}. 
Further optimization of the CVD parameters and the transfer procedure to 
produce solitary single-wall carbon nanotubes is thus required.

\section{Conclusions and outlook}

We have implemented a technique for carbon nanotube transfer separating growth 
and measurement onto different substrates. Nanotubes are grown on the tips 
of commercially available quartz tuning forks and subsequently transferred to a 
target chip of desired design. 

As with other nanotube transfer procedures, the  choice of contact materials and 
circuit elements for the target chip is much less constrained than for {\it in
situ} overgrowth, carbon nanotubes not suitable for measurements can easily be 
removed, and complex-structured devices can be re-used in more than one transfer 
attempt. Transfer targets may range from, e.g., superconducting coplanar circuit
geometries \cite{rmp-xiang-2013,aspelmeyer,pr-gu-2017}, qubit circuits 
\cite{prb-wang-2017}, superconducting single electron transistors 
\cite{prl-palyi-2012,prb-struck-2014}, or ferromagnetic contact electrodes 
\cite{prb-stadler-2015}, all the way to diamond crystallites containing 
NV-centers \cite{prl-li-2016}.

The quartz tuning forks are standardized, macroscopic parts that can be 
obtained in large numbers. In addition, they are highly robust, and survive 
multiple cycles of catalyst deposition, growth, nanotube transfer, and cleaning. 
This allows an easy, systematic approach towards integrating carbon nanotubes 
into devices of arbitrary complexity.

\section*{Acknowledgments}
\quad
The authors thank E. Weig for the initial suggestion of using quartz tuning 
forks, and Coftech GmbH for the quartz tuning fork wafer. Transport data has 
been recorded using the
\href{https://www.labmeasurement.de/}{Lab::Measurement} software package
\cite{labmeasurement}. We acknowledge funding by the Deutsche
Forschungsgemeinschaft via grants Hu 1808/1, SFB 689, and GRK 1570.

\bibliography{paper}

\end{document}